\documentclass[eqsecnum,showpacs,twocolumn]{revtex4}
\usepackage{xspace,amsfonts,amssymb,graphicx,color}
\begin{document}
\title{The Uncertainty Relation in ``Which-Way" Experiments: \\
How to Observe Directly  the Momentum Transfer using Weak Values}
\author{J. L. Garretson}
\affiliation{Centre for Quantum Dynamics, School of Science,
Griffith University,  Brisbane, 4111 Australia. }
\author{H. M. Wiseman}
\email{H.Wiseman@griffith.edu.au} \affiliation{Centre for Quantum
Dynamics, School of Science, Griffith University,  Brisbane, 4111
Australia. }
\author{D. T. Pope}
\affiliation{Centre for Quantum Dynamics, School of Science,
Griffith University,  Brisbane, 4111 Australia. }
\author{D. T. Pegg}
\affiliation{Centre for Quantum Dynamics, School of Science,
Griffith
University,  Brisbane, 4111 Australia. }

\newcommand{\beq}{\begin{equation}}
\newcommand{\eeq}{\end{equation}}
\newcommand{\bqa}{\begin{eqnarray}}
\newcommand{\eqa}{\end{eqnarray}}
\newcommand{\nn}{\nonumber}
\newcommand{\nl}[1]{\nn \\ && {#1}\,}
\newcommand{\erf}[1]{Eq.~(\ref{#1})}
\newcommand{\erfs}[2]{Eqs.~(\ref{#1})--(\ref{#2})}
\newcommand{\dg}{^\dagger}
\newcommand{\rt}[1]{\sqrt{#1}\,}
\newcommand{\smallfrac}[2]{\mbox{$\frac{#1}{#2}$}}
\newcommand{\half}{\smallfrac{1}{2}}
\newcommand{\bra}[1]{\langle{#1}|}
\newcommand{\ket}[1]{|{#1}\rangle}
\newcommand{\ip}[2]{\langle{#1}|{#2}\rangle}
\newcommand{\sch}{Schr\"odinger }
\newcommand{\schs}{Schr\"odinger's }
\newcommand{\hei}{Heisenberg}
\newcommand{\heis}{Heisenberg's }
\newcommand{\ito}{It\^o }
\newcommand{\str}{Stratonovich }
\newcommand{\dbd}[1]{{\partial}/{\partial {#1}}}
\newcommand{\sq}[1]{\left[ {#1} \right]}
\newcommand{\cu}[1]{\left\{ {#1} \right\}}
\newcommand{\ro}[1]{\left( {#1} \right)}
\newcommand{\an}[1]{\left\langle{#1}\right\rangle}
\newcommand{\st}[1]{\left|{#1}\right|}
\newcommand{\implies}{\Longrightarrow}
\newcommand{\del}{\nabla}
\newcommand{\du}{\partial}
\newcommand{\singlecol}{\end{multicols}
     \vspace{-0.5cm}\noindent\rule{0.5\textwidth}{0.4pt}\rule{0.4pt}
     {\baselineskip}\widetext }
\newcommand{\doublecol}{\noindent\hspace{0.5\textwidth}
     \rule{0.4pt}{\baselineskip}\rule[\baselineskip]
     {0.5\textwidth}{0.4pt}\vspace{-0.5cm}\begin{multicols}{2}\noindent}
\newcommand{\tick}{$\sqrt{\phantom{I_{I}\hspace{-2.1ex}}}$}
\newcommand{\cross}{$\times$}
\newcommand{\ww}{which-way }
\newcommand{\ps}[1]{\hspace{-5ex}{\phantom{\an{X_{w}}}}_{#1}\!}
\newcommand{\bbb}[1]{\hspace{-5ex}{\phantom{\an{X_{w}}}}_{#1}\!}
\newcommand{\mket}[1]{|\hspace{-0.3ex}\ket{#1}\hspace{-0.5ex}\rangle}
\newcommand{\mbra}[1]{\langle\hspace{-0.5ex}\bra{#1}\hspace{-0.3ex}|}

\newcommand{\red}[1]{\color{red}{#1}\color{black}}
\newcommand{\blu}[1]{\color{blue}{#1}\color{black}}

\begin{abstract}
A which-way measurement destroys the twin-slit interference
pattern. Bohr argued that this can be attributed to the \hei\
uncertainty relation: distinguishing between two slits a distance
$s$ apart gives the particle a random momentum transfer $\wp$ of
order $h/s$. This was accepted for more than 60 years, until
Scully, Englert and Walther (SEW) proposed a which-way scheme
that, they claimed, entailed no momentum transfer. Storey, Tan,
Collett and Walls (STCW) on the other hand proved a theorem that,
they claimed, showed that Bohr was right. This work reviews and
extends a recent proposal [Wiseman, Phys. Lett. A {\bf 311}, 285
(2003)] to resolve the issue using a {\em weak-valued} probability
distribution for momentum transfer, $P_{\rm wv}(\wp)$.  We show
that $P_{\rm wv}(\wp)$ must be nonzero for some $\wp: |\wp| >
h/6s$. However, its moments can be identically zero, such as in
the experiment proposed by SEW. This is possible because $P_{\rm
wv}(\wp)$ is not necessarily positive definite. Nevertheless, it
is measurable experimentally in a way understandable to a
classical physicist. The new results in this paper include the
following. We introduce a new measure of spread for $P_{\rm
wv}(\wp)$: half the length of the unit-confidence interval. We
conjecture that it is never less than $h/4s$, and find numerically
that it is approximately $h/1.59s$ for an idealized version of the
SEW scheme with infinitely narrow slits. For this example, the
moments of $P_{\rm wv}(\wp)$, and of the momentum distributions,
are undefined unless a process of apodization is used. However, we
show that by considering successively smoother initial wave
functions, successively more moments of both $P_{\rm wv}(\wp)$ and
the momentum distributions become defined. For this example the
moments of $P_{\rm wv}(\wp)$ are zero, and these moments are equal
to the changes in the moments of the momentum distribution. We
prove that this relation also holds for schemes in which the
moments of $P_{\rm wv}(\wp)$ are non-zero, but it holds {\em only
for the first two moments}. We also compare these moments to the
moments of two other momentum-transfer distributions that have
previously been considered, and with the moments of $\hat{p}_f -
\hat{p}_i$ (which is defined in the \hei\ picture).  We find
agreement between all of these, but again only for the first two
moments.  Our results reconcile the seemingly opposing views of
SEW and STCW.
\end{abstract}

\pacs{03.65.Ta$\\$Keywords:$\;$interference, momentum transfer,
measurement, slit}

\maketitle

\section{Introduction}

\subsection{History (to 1995)}

In a twin-slit experiment the far field interference pattern is a
picture of the transverse momentum distribution $P_{i}(p)$ of the
particle, with fringe spacing equal to $h/s$ (see Fig~1(a)).
\begin{figure}
\begin{center}
\includegraphics[width=.48\textwidth]{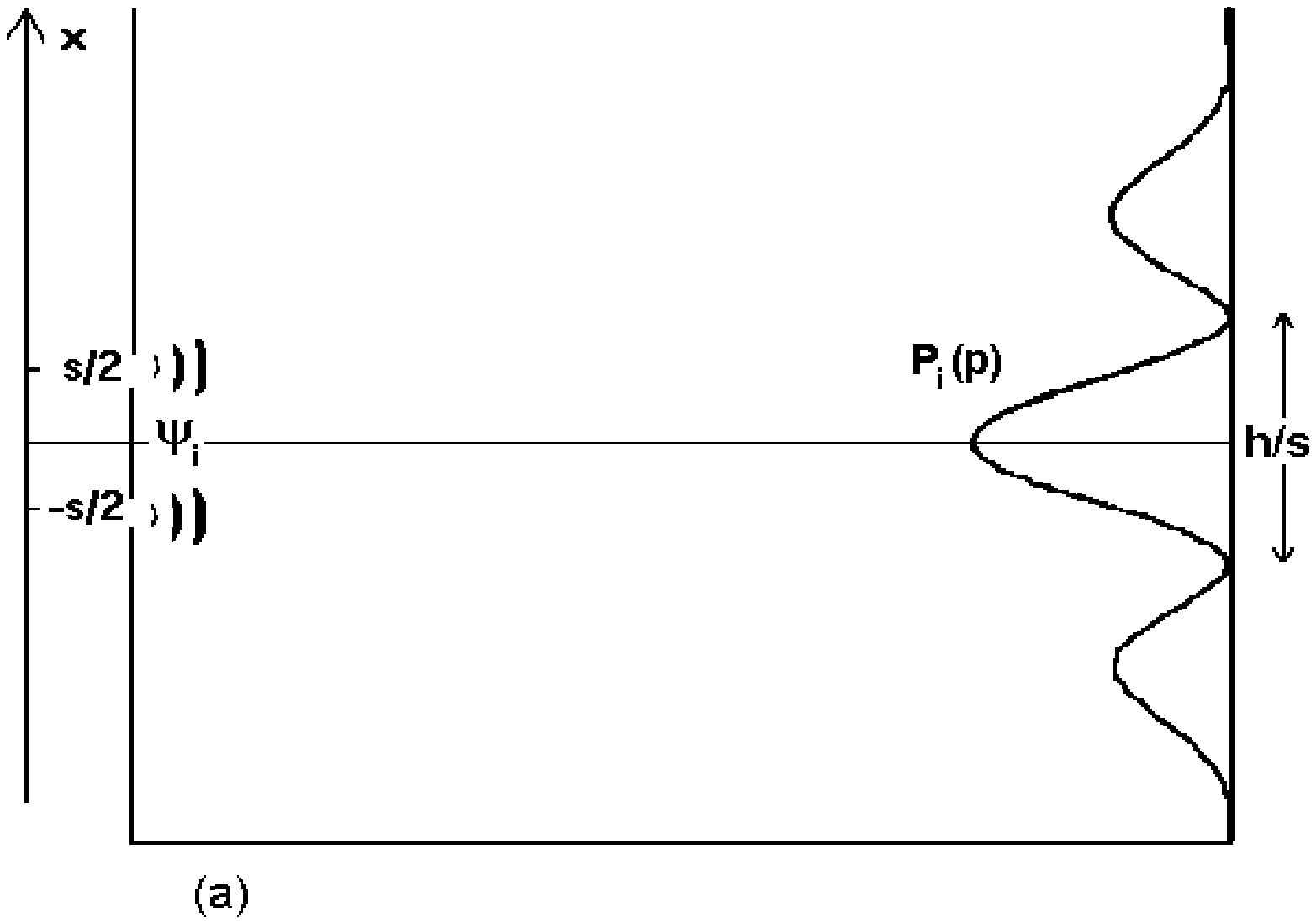}

\includegraphics[width=.48\textwidth]{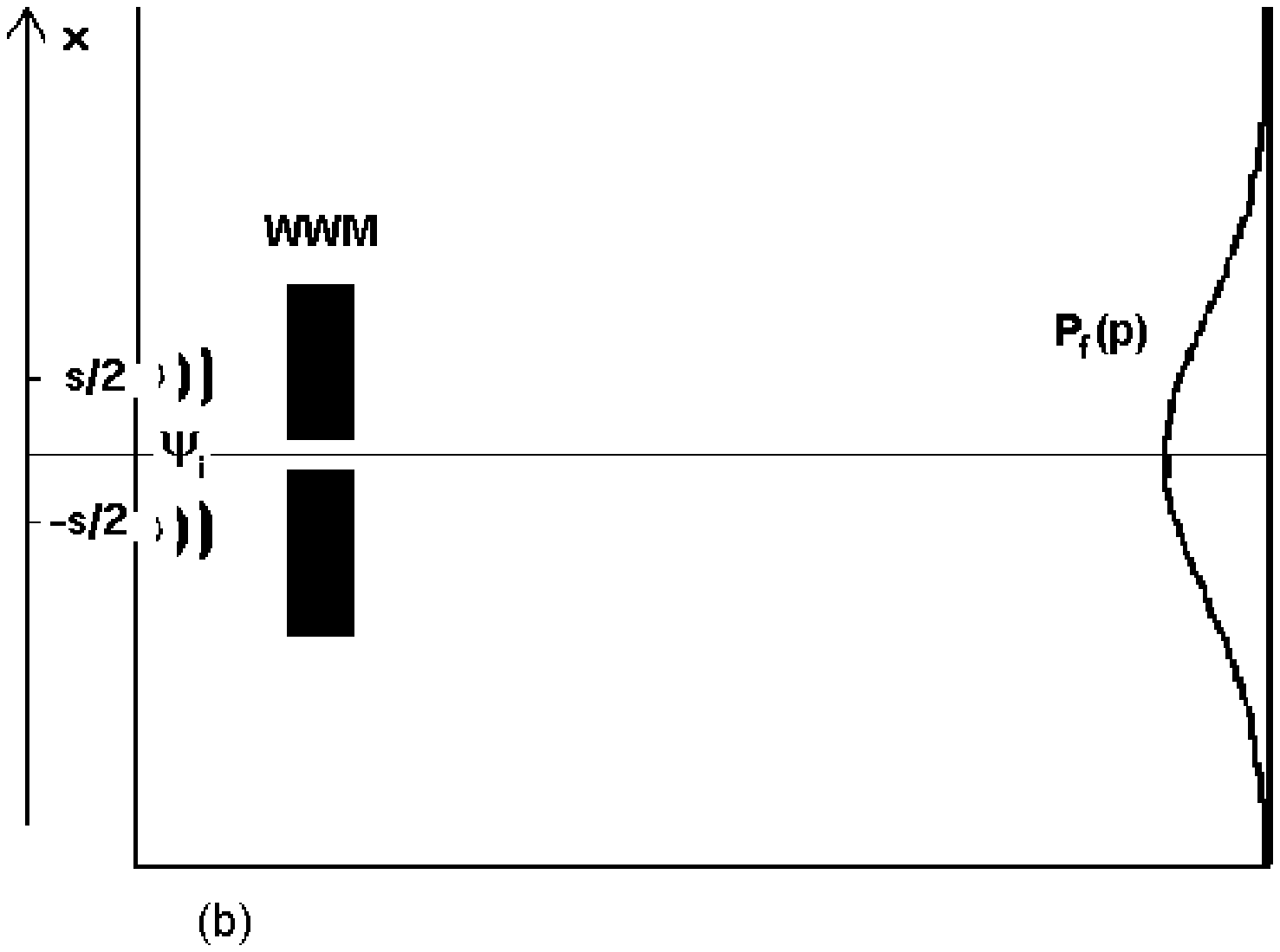}

\includegraphics[width=.48\textwidth]{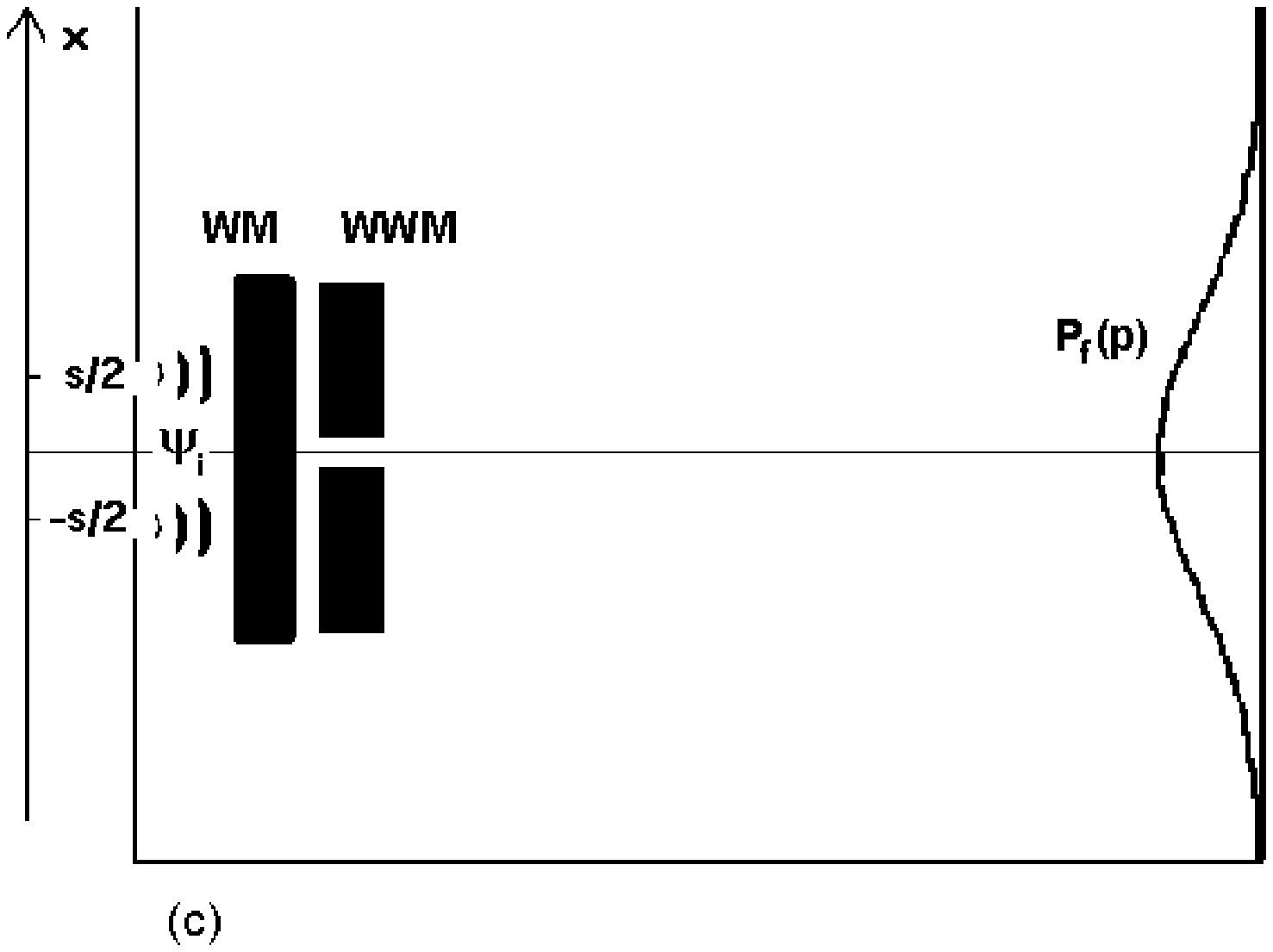}
\caption{Diagram of a twin-slit \ww experiment. The initial state
$\ket{\psi_{i}}$ is formed by the slits and propagates
longitudinally towards the final screen. This  is in the far
field, so detecting the position of the particle there is
equivalent to measuring its final momentum $p_{f}$. In (a) there
is no which-way measurement (WWM) device in place so $p_f = p_i$,
and its distribution is the twin-slit interference pattern. In (b)
the WWM device is in place and the distribution for $p_{f}$ is
just the single-slit diffraction pattern. In (c), a device is
added between the slits and the WWM device which makes a weak
measurement (WM) of $\ket{p_{i}}\bra{p_{i}}$, as explained later
in the text.}
\end{center}
\end{figure}
These fringes are a
signature of the particle being in a superposition of two
positions. Making a position measurement to determine through
which slit the particle passed changes the initial momentum
distribution to a final distribution $P_{f}(p)$ which lacks such
fringes (see Fig.~1(b)).  This is the canonical example of Bohr's
complementarity principle \cite{BohrEinst,WheZur83}.

To defend this principle against Einstein's recoiling slit {\em
gedankenexperiment}, Bohr relied upon the recently (in 1927)
derived Heisenberg uncertainty relation \cite{Hei27} to show that
the position measurement would cause an ``uncontrollable change in
the momentum'' $\wp \gtrsim h/s$, where $s$ is the slit separation
\cite{BohrEinst}. This is just what is required to wash out the
fringes in the momentum distribution, thereby enforcing
complementarity. Bohr's argument was famously reiterated by
Feynman \cite{FeyLeiSan65}, for a measurement using \heis light
microscope \cite{Hei27}.

Bohr's argument remained apparently unquestioned until 1991, when
Scully, Englert and Walther \cite{ScuEngWal91} proposed a new
which-way (or {\em welcher Weg}) measurement (WWM) for which, they
calculated, no momentum would be transferred to the particle. Thus
they concluded that the arguments of Feynman and Bohr were wrong
in general, and that complementarity must be deeper than
uncertainty. Their calculation consisted in showing that a {\em
single-slit} wavefunction was unchanged by their WWM. This is in
contrast to earlier WWMs, such as considered by Feynman and Bohr,
which cause an increase in the variance of the single-slit
momentum distribution (the diffraction pattern) by of order
$(\hbar/s)^2$.

Moreover, it can be shown \cite{Wis97a} that this new feature of
the SEW scheme translates into a quantitative difference in the
{\em twin-slit} pattern. In earlier schemes the variance of the
final distribution is greater than that of the initial one,
because
\beq \label{convprob} P_{f}(p) = \int d\wp P_{\rm
cl}(\wp) P_{i}(p-\wp). \eeq
That is, the momentum disturbance can be treated as a
classical mixture of mutually exclusive momentum kicks.
The probability distribution  $P_{\rm cl}(\wp)$ for transferring momentum $\wp$
is a true (i.e. positive) probability distribution.
In the scheme of SEW (and other schemes discussed in Ref.~\cite{Wis97a}),
the relation (\ref{convprob}) does not hold. That is, the effect
of the SEW WWM is not equivalent to giving the particle  a
classical momentum kick $\wp$ chosen randomly from a distribution
$P_{\rm cl}(\wp)$.

For `classical' WWM schemes, for which \erf{convprob} applies, it
follows that the differences in the mean and variance of
$P_{f}(p)$ and $P_{i}(p)$ are equal to the mean and variance of
$P_{\rm cl}(\wp)$. Since the width of $P_{\rm cl}(\wp)$ must be of
order $\hbar/s$ to wash out the fringes, it follows that in such
cases the variance of $P_{f}(p)$ must be greater than that of
$P_{i}(p)$ by of order $(\hbar/s)^2$. For nonclassical schemes,
such as that of SEW, such a proof does not hold. This difference
was first pointed out in Ref.~\cite{WisHar95}. In particular, for
the SEW scheme, the mean and variance of the momentum distribution
are unchanged by the WWM \cite{Wis97a}.

The argument of SEW was not accepted by Storey, Tan, Collett and
Walls (STCW)  \cite{StoTanColWal94}. Since $P_{i}(p)$ has fringes,
but $P_{f}(p)$ does not, they reasoned that the momentum must have
been disturbed. To back this up, they  proved a theorem that
showed that if interference is destroyed then there is always some
transverse momentum transfer $\geq \hbar/s$. This lower bound of
$\hbar/s$ was strengthened in Ref.~\cite{Wis97a} to $\pi\hbar/2s =
h/4s$. It was also shown there that the theorem of STCW has the
following experimentally verifiable consequence.  If the particle
were prepared in a momentum eigenstate, with $P_{i}(p)  =
\delta(p)$, then under any WWM the final momentum distribution
would be such that \beq P_{f}(p) \neq 0 \textrm{ for some } p
\textrm{ such that } |p| > h/4s. \eeq

Since \erf{convprob} does not always hold, STCW found their
momentum transfer not in a probability distribution $P_{\rm
cl}(\wp)$, but in a set of probability {\em amplitude}
distributions $\tilde{O}_\xi(\wp)$. The debate
\cite{QI94,Nature95} between SEW and STCW did not lead to any
progress in understanding because neither side appeared to
appreciate that the other side was using a different {\em concept}
of momentum transfer. It was only in Ref.~\cite{WisHar95} that
this distinction was clearly made, so that it could be seen that
each side had a valid point of view.

\subsection{Concepts of Momentum Transfer}

The reader might be forgiven for wondering why there is such
difficulty with the simple-sounding concept of momentum transfer.
Classically, one would just measure the initial momentum $p_{i}$
(just after the slits) and then the final momentum $p_{f}$ (after
the WWM), and determine the distribution for $\wp \equiv
p_{f}-p_{i}$. The problem with this procedure quantally is that
measuring $\hat{p}_i$ precisely destroys the twin-slit
wavefunction --- it creates a momentum eigenstate.

 If one were to follow this procedure then one would
 find that $\wp = p_{f}-p_{i}$ has the
same distribution considered above in the context of the argument
of STCW. That is, one would find that the absolute value of the
momentum transfer $\wp$ would sometimes be larger than $h/4s$. For
cases of classical momentum kicks, the distribution for $\wp$ is
precisely the $P_{\rm cl}(\wp)$ in \erf{convprob}, but even
in nonclassical cases the width of the distribution must be larger
than $h/4s$.

On the other hand, one could well argue that the effect of the WWM
on a momentum eigenstate (which is spread over all position) is
not the same as its effect on the twin-slit wavefunction. But
since this is disturbed by a momentum measurement, one seems
restricted to considering the change in the momentum distribution
from $P_{i}(p)$ before (or without) the WWM to $P_{f}(p)$ after
(or with) the WWM. These are clearly different (the latter lacks
fringes), but in terms of the moments, one would find the result
quoted above in the context of the argument of SEW. That is, one
would find that the mean and variance of the two distributions are
identical for the SEW scheme.

One could simply accept that there are two concepts of momentum
transfer, and they cannot be reconciled in general. However, it is
tempting to try to find a formalism which can get around the
above-mentioned problem of the disturbance induced by a
measurement of $p_{i}$.  What is required is a quantum formalism
which somehow treats $p_i$ as being definite even when it is not
known via a precise measurement.

Two such formalisms have been investigated in the past by one of
us and co-workers. The first is the  Wigner function
\cite{Wis97a}. The second is Bohmian mechanics \cite{Wis98a}.
Unlike the original works by SEW and STCW, each of these
formalisms completely characterizes the momentum transfer $\wp$ by
giving a distribution for it. In each case this distribution also
depends upon another variable, the transverse position of the
particle in the Wigner case, and the time (or longitudinal
position) after passing through the WWM in the Bohmian case.
Interestingly, in each case the distributions  reflect both the
results of SEW and of STCW. The interested reader is referred to
the original works and the comparison in Ref.~\cite{Wis03}.

The main drawback of these formalisms is that they are too formal.
That is, there is no way to observe directly the distributions of
momentum transfer they generate. By `to observe directly' we mean to
obtain these distributions from an experiment in a way that would
be completely understandable to a classical physicist.

This is the motivation for the current approach, first introduced
in Ref.~\cite{Wis03},  of using the {\em weak value} theory of
Aharanov, Albert, and Vaidman \cite{AhaAlbVai88}. In a nutshell,
the idea is to obtain information about the initial momentum by
making a {\em weak} measurement, so as to disturb the initial
wavefunction only negligibly. As will be shown, this approach
enables one to observe directly  a weak-valued probability
distribution $P_{\rm wv}(\wp)$ for momentum transfer $\wp$.
Moreover, this distribution reflects both the position of SEW and
that of STCW. This surely is the best resolution of the debate
that could possibly be hoped for.

\subsection{Organization of this Paper}

The remainder of this paper is organized as follows. In Sec.~II we
summarize the theory of weak values as introduced by Aharonov,
Albert and Vaidman, and motivate their application to the current
issue. In Sec.~III we review the formal description of WWMs. Note
that we do this in a way different from that adopted previously,
including in Ref.~\cite{Wis03}, in order to integrate weak values
into the theory in a more natural way. In Sec.~IV we do just that,
deriving the expression for $P_{\rm wv}(\wp)$ more elegantly than
in Ref.~\cite{Wis03}.

In Sec.~V we expand upon the (very brief) derivations in
\cite{Wis03} of the elementary properties of $P_{\rm wv}(\wp)$
which show how it is compatible with both SEW and STCW. In Sec.~VI
we illustrate the properties of $P_{\rm wv}(\wp)$ using a number
of different examples.  For the simplest conceivable measurement,
we calculate $P_{\rm wv}(\wp)$ for a number of different initial
states.  We also show that the moments of $P_{\rm wv}(\wp)$ (when
they are defined) are equal to zero, and equal to the change in
the moments from the initial $P_i(p)$ to the final $P_f(p)$
momentum distributions of the particle. In Sec.~VII we calculate
the first three moments of the momentum transfer distributions
with general WWMs for a number of different momentum-transfer
formalisms and show that in general the first and second moments
of all of these distributions is equal to the change in the
moments of $P_f(p)$ and $P_i(p)$. However this relationship ceases
to hold for higher order moments for all of these formalisms. We
conclude in Sec.~VIII.

\section{Weak Values}

\subsection{Introduction}

It is a fundamental fact of quantum theory that a projective
measurement, which one could also call a  precise or strong
measurement,  greatly disturbs the quantum state  in general.
However one can consider imprecise measurements (which are
non-projective), for which the disturbance {\em can be} small.
(The disturbance is not necessarily small because the imprecision
may be due simply to poor control of classical noise). A {\em weak
measurement} of a quantity is one which is arbitrarily imprecise,
and for which the disturbance is correspondingly arbitrarily
small.

A {\em weak value} is just the  {\em mean} value of a weak
measurement. That is, it is obtained by averaging over a large
ensemble of weak measurement results on identically prepared
systems, just as is the mean value of a strong measurement.
However, because of the imprecision in each weak measurement
result, the size of the ensemble must be correspondingly larger
than in the case of strong measurements.

Simply considering a prepared state $\ket{\psi}$ gives an
uninteresting weak value
 --- the same as the strong value for the same quantity:
\beq
 \an{X_{\rm weak}}_{\psi} = \an{X_{\rm strong}}_{\psi} =
\bra{\psi}\hat{X}\ket{\psi}.
 \eeq
As realized by Aharonov, Albert, and Vaidman \cite{AhaAlbVai88},
to obtain an interesting weak value requires {\em post-selection}.
That is, the average is calculated from the sub-ensemble where a
{\em later} strong measurement  reveals the state to be
$\ket{\phi}$.

Allowing for some evolution $\hat{U}$ after the weak measurement,
the post-selected weak value turns out to be \beq
\label{weakvalgen} \bbb{\phi}\an{X_{\rm w}}_{\psi} = {\rm
Re}\frac{\bra{\phi}\hat U\hat{X}\ket{\psi}}{\bra{\phi}\hat
U\ket{\psi}}. \eeq The interested reader is referred to the
appendix for a very brief outline of how this formula may be
derived.

This expression is certainly unusual, in that the numerator and
denominator are linear in $\ket{\psi}$ and $\ket{\phi}$ rather
than bilinear. This has the consequence that the weak value can
lie {\em outside} the range of eigenvalues of $\hat{X}$
\cite{AhaAlbVai88}. This was soon verified experimentally
\cite{RitStoHul91}. This of course cannot happen for a  strong
measurement of $\hat{X}$, for which the post-selected strong value
would be \beq \bbb{\phi}\an{X_{\rm s}}_{\psi} = \frac{\sum_{x}
\st{\bra{\phi}\hat U\ket{x}}^{2} x\st{\ip{x}{\psi}}^{2}}
{\sum_{x'}  \st{\bra{\phi}\hat U\ket{x'}}^{2}
\st{\ip{x'}{\psi}}^{2}}. \eeq

\subsection{The Motivation}

There are three motivations for considering weak values in the
context of momentum transfer in  WWMs.

The first is that weak values have a good record for supplying new
insight into quantum puzzles. They have been used to define
tunneling time in a directly observable manner \cite{Ste95} and to
resolve Hardy's paradox \cite{Aha01}. With a few simple
generalizations, weak values have also been found \cite{Wis02a} to
explain ``anticausal" conditional quadrature evolution in a
well-known cavity QED experiment \cite{FosOroCasCar00}.

The second motivation is more specific. To investigate momentum
transfer in WWMs one wants to know $\wp = p_{f} - p_{i}$ without
making a strong measurement of $p_{i}$. It is an obvious (in
hindsight) idea to make a weak measurement.

Thirdly, weak values will enable experimental investigation of the
problem, by direct observation of the momentum transfer. The lack
of meaningful predictions that are interesting enough to be tested
by experimentalists is, in our opinion, one of the problems with
this area of research. We note in passing that the experiment by
Rempe and co-workers \cite{DurNonRem98}, as interesting and
elegant as it was, was not relevant to the debate between SEW and
STCW. This is simply because it did not involve a which-way
measurement of a particle prepared in a superposition of two
positions.

\section{The WWM formalism}

As noted previously, we are concerned with the case where the
quantum particle is prepared initially in a state $\ket{\psi_{i}}$
which is a superposition of two states that are well-localized in
position (that is, at the two slits, centred at $\pm s/2$). It is
necessary to restrict the discussion to an interferometer of this
kind, where the initial superposition is in transverse position
and the free evolution preserves the conjugate quantity
(transverse momentum), so that the issue of loss of visibility
relates directly to the transverse momentum transfer.

In the WWM the particle is coupled to a meter, whose state we will
distinguish from that of the system by using a double ket. For
instance, the initial state of the meter is $\mket{\theta}$. The
coupling of the  system and meter can be described by a unitary
operator $\hat{U}_{{}_{\rm WWM}}$. This enables the meter to
obtain which-way information and as a consequence also change the
momentum distribution of the particle. The \ww information can be
obtained by the experimenter by reading out the meter. For
simplicity we will assume that this is performed by making a
measurement of the meter in a complete orthogonal basis
$\{\mket{\xi}\}$. Thus the evolution of the system and meter is as
follows: \bqa \mket{\theta}\ket{\psi_{i}} &\to& \hat{U}_{{}_{\rm
WWM}}\mket{\theta} \ket{\psi_{i}} \\ &\to&  \mket{\xi}\mbra{\xi}
\hat{U}_{{}_{\rm WWM}}\mket{\theta}\ket{\psi_{i}} \, = \,
\mket{\xi}\hat{O}_\xi \ket{\psi_{i}}. \label{finalstate} \eqa

In \erf{finalstate},  $\hat{O}_\xi \equiv \mbra{\xi}
\hat{U}_{{}_{\rm WWM}}\mket{\theta}$ is an operator in the system
Hilbert space called the {\em measurement operator}. The WWM is
completely described by the set of measurement operators $\{\hat
O_\xi\}$. The probability to obtain the result $\xi$ is
$\bra{\psi_{i}}\hat{O}_\xi\dg \hat O_\xi\ket{\psi_{i}}$. The final
state in \erf{finalstate} must be divided by the square root of
this in order for it to be normalized.

For a WWM we need to obtain information about the position of the
particle. That means that we want the measurement operators to be
functions of the position operator:  $\hat{O}_\xi = O_\xi(\hat
x)$. Thus the WWM is described by the set of functions
$\cu{O_\xi(x)}$, which are restricted only by the completeness
relation \beq \label{complet} \sum_\xi |O_\xi(x)|^2 =
1 \; \forall x . \eeq This is necessary for the probabilities of the results
$\xi$ to sum to unity.

For some WWMs, the read-out basis $\mket{\xi}$ can be chosen such
that ${O}_{\xi} (\hat x) =
\sqrt{N_{\xi}}\exp(-ik_{\xi}\hat{x})$ for all $\xi$, where obviously $\sum_\xi
N_\xi = 1$. In such cases \erf{convprob} can be shown to pertain
\cite{WisHar95,Wis97a}, where \beq \label{PclassfromO} P_{\rm
cl}(\wp) = \sum_{\xi} N_{\xi}\delta(\wp-\hbar k_{\xi}) \eeq
That is, $N_\xi$ can be interpreted as the probability for the
system to receive a momentum kick equal to $\hbar k_\xi$.

In general, \erf{convprob} does not pertain, but an analogous
equation for probability amplitudes does, \beq \label{convmom}
\bra{p}\hat{O}_{\xi}\ket{\psi_{i}}
 = \int\! d\wp\, \tilde{O}_{\xi}(\wp) \tilde\psi_{i}(p-\wp).
\eeq Here, \beq \label{fourier} \tilde{O}_{\xi}(p) = (2\pi
\hbar)^{-1/2}\int\! dx\, O_{\xi}(x) e^{-ix p/\hbar} \eeq is the
probability amplitude for a momentum kick $\wp$, as identified by
STCW\cite{StoTanColWal94} and similarly \beq \label{initmom}
\tilde{\psi}_i(p) = (2\pi \hbar)^{-1/2}\int\! dx\, \psi_i(x)
e^{-ix p/\hbar}, \eeq where $\psi_i(x) = \bra{x}{\psi_i}\rangle$.

For narrow slits, [i.e. $|\psi(x)|^{2} \simeq
\delta(2x+s)+\delta(2x-s)$], the visibility of the far field
interference pattern can be shown \cite{Wis97a} to be given by
\beq \label{vis} V = \st{ \sum_\xi O_{\xi}(-s/2)O_{\xi}^{*}(s/2)}.
\eeq From this and \erf{convmom}, it can be shown \cite{Wis97a}
that the WWM will disturb a momentum eigenstate by an amount at
least equal to $\arccos(V)\hbar/s$. For $V=0$ one obtains the
previously stated lower bound, $h/4s$. Since the visibility is
only defined for a twin-slit wavefunction, the disturbance to a
momentum eigenstate actually reflects the {\em quality} of the WWM
(which exists independently of the visibility of the fringe
pattern). This quantity, $Q$, is defined and analysed in
Ref.~\cite{MarHar03}.

\section{Applying Weak Values to WWMs}

Clearly to apply weak values to WWMs it is necessary to make the
weak measurement on $\ket{\psi_{i}}$ before the WWM, followed by a
strong measurement of $\hat{p}$ (which is simply a measurement of
position in the far-field), as shown in Fig.~1(c).

A first thought would be to make a weak measurement of $\hat{p}$.
Including the initial meter state and the final meter state (for a
particular result $\xi$), and describing the WWM by the unitary
operator $\hat U_{{}_{\rm WWM}}$, one can apply \erf{weakvalgen}.
This yields the weak value
\beq \label{notwww}
\bbb{p_{f},\xi}\an{p_{\rm w}}_{\theta,\psi_{i}} =  {\rm
Re}\frac{\bra{p_{f}}\mbra{\xi} \hat U_{{}_{\rm WWM}}\hat{p}\,
\mket{\theta}\ket{\psi_{i}}}{\bra{p_{f}}\mbra{\xi}\hat U_{{}_{\rm
WWM}}\mket{\theta}\ket{\psi_{i}}}. \eeq That is, one could measure
the weak value of the initial momentum $p_{i}$, post-selected on
the final momentum $p_{f}$ and WWM result $\xi$.

A little consideration of \erf{notwww} reveals that this is not
what we want. Say it turned out that the weak value of the initial
momentum was the same as the final momentum:
$\bbb{p_{f},\xi}\an{\hat{p}_w}_{\theta,\psi_{i}}  = p_{f}$. Then
all that one could say would be that the WWM (for result $\xi$)
does not change  the {\em mean} momentum. This will be the case
for any symmetric disturbance of the momentum. To address the
momentum transfer issue we need to know of {\em any} disturbance
to the momentum.

A second (and better) thought is to make a weak measurement of the
{\em projector} $\hat{\pi}(p_{i}) \equiv \ket{p_{i}}\bra{p_{i}}$
for some particular $p_i$. The non-post-selected mean value of a
measurement (weak or strong) of $\hat\pi(p_{i})$ on
$\ket{\psi_{i}}$ would give the {\em probability} $P_{i}(p_{i}) =
\ip{\psi_i}{p_i}\!\ip{p_{i}}{\psi_{i}}$. The post-selected weak
value \beq \label{one} \bbb{\xi,p_{f}}\an{\pi(p_{i})_{\rm
w}}_{\theta,\psi_{i}} =  {\rm Re}\frac{\bra{p_{f}}\mbra{\xi}\hat
U_{{}_{\rm WWM}} \hat{\pi}(p_{i})\,\mket{\theta}\ket{\psi_{i}}}
{\bra{p_{f}}\mbra{\xi}\hat U_{{}_{\rm WWM}}
\mket{\theta}\ket{\psi_{i}}} \eeq can thus be interpreted as the
weak-valued {\em conditional probability} for the initial momentum
being $p_i$, given $p_f$ and $\xi$. We will denote this $P_{\rm
wv}(p_{i}|\xi,p_{f})$.

Using this expression for $P_{\rm wv}(p_{i}|\xi,p_{f})$, we can
define a weak-valued joint probability distribution \beq
\label{two} P_{\rm wv}(p_{i};\xi,p_{f}) = P_{\rm
wv}(p_{i}|\xi,p_{f}) \times P(\xi,p_{f}), \eeq where \beq
\label{three} P(\xi,p_{f}) = \st{\bra{p_{f}}\mbra{\xi}\hat
U_{{}_{\rm WWM}} \mket{\theta}\ket{\psi_{i}}}^2 \eeq is the
probability to obtain the result $\xi$ and the final momentum
$p_f$.

We are now almost at our goal of quantifying the momentum transfer
$\wp = p_{f}-p_{i}$. We rewrite $P_{\rm wv}(p_{i};\xi,p_{f})$ in
terms of $\wp$, $p_i$, and $\xi$, and then average over all
$p_{i}$, and over all results $\xi$. This means repeating the
experiment many times for all choices of $p_i$. The result is the
weak-valued probability distribution for the momentum transfer
$\wp$, \beq P_{\rm wv}(\wp) \equiv \sum_\xi \int dp_{i} P_{\rm
wv}(p_{i};\xi,p_{i}+\wp) \eeq

Using \erfs{one}{three} and the definition of $\hat{O}_\xi$,
one finds that $P_{\rm wv}(\wp)$ is equal to
\beq \sum_\xi
\int\! dp_{i}\, {\rm Re} 
\cu{\bra{p_{i}+\wp}\hat{O}_{\xi}
\ket{p_{i}}\ip{p_{i}}{\psi_{i}}\bra{\psi_{i}}\hat{O}_{\xi}\dg\ket{p_{i}+\wp}}.
\label{wvmt1} \eeq

Finally, using $\hat{O}_\xi = O_\xi(\hat x)$, one can
straightforwardly derive the remarkably simple and elegant formula
\beq \label{wvmt2} P_{\rm wv}(\wp) = \sum_\xi {\rm Re}\cu{
\tilde{O}_{\xi}(\wp) \tilde{Q}_{\xi}^{*}(\wp)}. \eeq The Fourier
transform is as defined in \erf{fourier}, and \beq Q_{\xi}(x) =
O_{\xi}(x) |\psi_{i}(x)|^{2}. \eeq Thus the weak-valued
probability
distribution for $\wp$ depends upon the initial state in a very
natural way.

\section{Elementary Properties of $P_{\rm wv}(\wp)$} \label{sec:elemprops}

\subsection{Proofs}

A number of interesting properties of $P_{\rm wv}(\wp)$ can now easily be proven.

First,  $P_{\rm wv}(\wp)$ is normalized. This is easily seen
 using the
moment-generating function
\bqa
\Phi(q) &\equiv& \int \!d\wp\, P_{\rm wv}(\wp) e^{i\wp q/\hbar} \\
&=& \sum_\xi  \int \!dx\, |\psi_{i}(x)|^{2} \half \bigl[
O_{\xi}(x)O_{\xi}^{*}(x-q) \nl{\phantom{\sum_\xi  \int \!dx\,} +}
O_\xi^*(x) O_\xi(x+q) \bigr] . \label{mgf} \eqa
Note that this
expression [and \erf{Phiforslits}] correct a mistake in
Ref.~\cite{Wis03} which did not affect the results stated there.
 As we will see later,
$P_{\rm wv}(\wp)$ may take negative values,  but it integrates to
\beq \Phi(0)= \int dx \sum_\xi  |O_\xi(x)|^2 |\psi_{i}(x)|^2 =
\int dx  |\psi_{i}(x)|^2 = 1. \eeq Moreover, from \erf{mgf} it can
be shown that $|\Phi(q)|\leq 1$, as it would be for a true
probability distribution. This is the case as \bqa
\label{limitonphi1} |\Phi(q)| &\leq& \sum_\xi\int dx \frac12
\left.|O_\xi^{*}(x-q)O_\xi(x)\right.\nn\\&& \phantom{\sum_\xi\int
dx \frac12} \left. +O_\xi^{*}(x)O_\xi(x+q)|
\,|\psi_{i}(x)|^{2} \right.\nn \\
&\leq& \sum_\xi\int dx \frac12 \left. \bigl[
|O_\xi^{*}(x-q)O_\xi(x)|\right.\nn\\ && \phantom{\sum_\xi\int dx
\frac12} \left. +|O_\xi^{*}(x)O_\xi(x+q)|\bigr]
\,|\psi_{i}(x)|^{2}, \right. \nn\\\eqa which follows from the
triangle inequality.  Using the Schwarz inequality we obtain \bqa
|\Phi(q)| &\leq& \sum_\xi\int dx \frac12 \left.
\bigl[ |O_\xi(x-q)||O_\xi(x)|\right.\nn\\
&& \phantom{\sum_\xi\int dx \frac12}\left.
+|O_\xi(x)||O_\xi(x+q)|\bigr] |\psi_{i}(x)|^{2}\right.\nn
\\
&\leq& \sum_\xi\int dx \frac14 \left.
\bigl[ |O_\xi(x-q)|^{2}+2|O_\xi(x)|^{2}\right. \nn\\
&& \phantom{\sum_\xi\int dx \frac12}\left.+|O_\xi(x+q)|^{2}\bigr]
|\psi_{i}(x)|^{2}, \right. \label{limitonPhi2} \eqa where the last
line follows from the fact that $(|A|-|B|)^2 \geq 0$. Using the
completeness relation from \erf{complet}, then, we have achieved
$|\Phi(q)| \leq 1$, as desired.

Second, in the case of a classical momentum disturbance
(\ref{PclassfromO}), it is easy to see that \beq P_{\rm wv}(\wp) =
P_{\rm cl}(\wp), \eeq which is a true probability
distribution. This is an important test-case. It shows that if
$P_{\rm wv}(\wp)$ takes negative values for some WWM scheme, that
scheme must involve a nonclassical momentum disturbance. As a
special case, if there is no WWM at all, $P_{\rm wv}(\wp) =
\delta(\wp)$, indicating that there is no momentum disturbance, as
expected.

Third, since the moments of $\wp$ are given by \beq
\an{\wp^{n}}_{\rm wv} = \ro{-i\hbar \dbd{q}}^{n}\Phi(q)|_{q=0},
\eeq it follows from \erf{mgf} that if the $O_{\xi}$ are flat
(i.e. have all derivatives zero) in the region of the slits where
$|\psi_{i}(x)|^{2}$ is nonzero, then all of the moments of
$P_{\rm{wv}}(\wp)$ are zero. This is the case (to a very good
approximation) in the scheme of SEW. Thus the claim that their
scheme would not transfer any momentum to the particle could be
validated experimentally by calculating the moments of the
measured $P_{\rm wv}(\wp)$.

Fourth and finally, despite this last fact, $P_{\rm wv}(\wp)$ also
reflects the change in the momentum distribution caused by a WWM,
as we now show. For narrow slits at $x=\pm s/2$, \beq
\label{Phiforslits} \Phi(s)= \frac14 \left[ {\cal
V}\left(\frac{3s}{2},\frac{s}{2}\right)  + 2{\cal
V}\left(\frac{s}{2},\frac{-s}{2}\right) + {\cal
V}\left(\frac{-s}{2},\frac{-3s}{2}\right) \right], \eeq where \beq
\label{defcalV} {\cal V}(x,x') \equiv \sum_\xi
O_\xi(x)O_\xi^{*}(x'). \eeq From \erf{vis}, the visibility $V$ of
the interference pattern is $|{\cal
V}(\frac{s}{2},\frac{-s}{2})|$. With a WWM in place, this will be
zero. Hence, by the triangle inequality, we have
 \bqa
\label{absphiforslits2} |\Phi(s)| &\leq& \frac14 \sq{ \st{{\cal
V}\ro{\frac{3s}{2},\frac{s}{2}}}+\st{{\cal
V}\ro{\frac{-s}{2},\frac{-3s}{2})}}} \nn  \\
&\leq& \frac12 . \eqa Here we have used $|{\cal V}(x,x')| \leq 1$,
which follows from the definition in \erf{defcalV}. Since in
addition $\Phi(0) = 1$ and $|\Phi(q)| \leq 1\;\forall q$, it
follows from the theorem in Appendix A of Ref.~\cite{Wis97a} that
\beq {\rm Support}[P_{\rm wv}(\wp)] \not\subset (-h/6s,h/6s).
\label{support} \eeq That is, $P_{\rm wv}(\wp)$ must be nonzero
for some $\wp$ having a magnitude at least equal to $h/6s$. This
supports the view of STCW. For an imperfect WWM where $V \neq 0$,
$h/6s$ must be replaced by $(\hbar/s) \arccos[ (V+1)/2 ]$
\cite{fn1}.

\subsection{Conjectures}

In Ref.~\cite{Wis97a} a similar lower bound to that in
\erf{support} was proven for (a) the final momentum distribution
where the system had been prepared in the zero-momentum
eigenstate, and (b) the ``nonlocal'' momentum transfer function in
the Wigner representation.  In these cases, the lower bound was
greater: $h/4s$ rather than $h/6s$. Also, in these cases, the
lower bound could be achieved by using a classical momentum
disturbance with \beq \label{mindistWig} P_{\rm cl}(\wp) =
\frac{1}{2} \left[ \delta\ro{\wp+\frac{h}{4s}} +
\delta\ro{\wp-\frac{h}{4s}} \right]. \eeq Because of the
similarity of the analysis in the cases of Ref.~\cite{Wis97a} and
the weak-valued momentum transfer probability distribution here,
we conjecture that the lower bound of $h/6s$ in \erf{support}
cannot be achieved. Rather, we conjecture, that the achievable
lower bound is $h/4s$, and that it is achieved for $P_{\rm
wv}(\wp) = P_{\rm cl}(\wp)$ in \erf{mindistWig}.

Although the measure \beq [\delta \wp]_{\rm support}  = {\rm min}
\cu{p: \;{\rm Support}[P_{\rm wv}(\wp)] \subset  (-p,p)} \eeq is
useful in that we can prove a lower bound of $h/6s$ and conjecture
a lower bound of $h/4s$, its disadvantage  as a measure of the
width of $P_{\rm wv}(\wp)$ is that for nonclassical measurements
it is typically infinite, as we will see in Sec.~VI. For classical
distributions, this measure is the same as the $\infty$-norm of
the distribution, where the $n$-norm is defined as \beq [\delta
\wp]_{n-{\rm norm}} = \sq{ \int d\wp P_{\rm cl}(\wp)
|\wp|^n}^{1/n}. \label{n-norm}\eeq However, in general this may be
undefined even for $n=1$, unless the distribution is apodized, as
discussed in Sec.~\ref{sec:unbdd}.

For classical distributions, $[\delta \wp]_{\infty-{\rm norm}}$ is
also equivalent to the unit-confidence interval width, $[\delta
\wp]_{1-{\rm confidence}}$, where \beq [\delta \wp]_{\epsilon-{\rm
confidence}} =  p : \int_{-p}^p P_{\rm wv}(\wp) d\wp = \epsilon
\eeq For non-positive distributions this definition still applies,
with a slight modification: \beq [\delta \wp]_{\epsilon-{\rm
confidence}} =  {\rm min}\cu{p : \int_{-p}^p P_{\rm wv}(\wp) d\wp
= \epsilon} \eeq The interesting point with non-positive
distributions is that $[\delta \wp]_{1-{\rm confidence}}$ may be
finite, even if $[\delta \wp]_{\rm support}$ is infinite.

We will see  that $[\delta \wp]_{1-{\rm confidence}}$ is a good
measure of the width for the very  nonclassical $P_{\rm wv}(\wp)$
in Sec.~\ref{sec:unbdd}. Moreover,  it evaluates to a quantity of
order, but larger than $h/4s$. 
On this basis, we
conjecture, as we did for $[\delta \wp]_{\rm support}$, that this
measure satisfies \beq [\delta \wp]_{1-{\rm confidence}} \geq
\frac{h}{4s}, \eeq and that the lower bound is met only for the
classical distribution (\ref{mindistWig}).

\section{Examples}

The results of the previous section can be illustrated by a number
of examples with different initial wavefunctions $\psi_i(x)$.
These examples all use a minimal WWM that only distinguishes
between $x<0$ and $x>0$.  That is, $O_\pm(x)=\Theta(\pm{x})$, the
Heaviside function.  This is an idealization of the WWM of SEW. In
this instance, of course $O_\pm(x)$ are perfectly flat in the
region of the slits, so from the above arguments all moments
should vanish in each of these cases.

The simplest example of infinitely narrow slits considered in
\cite{Wis03} proves to have difficulty in showing some of the nice
properties of $P_{\rm{wv}}(\wp)$.  This is because all moments
above zero order are undefined in a strict sense.  The theory of
apodization \cite{Zema} is required in order to give the
``correct" moments in this case.  For better behaved initial
states, however, this becomes less of a problem, as we shall see.
For an arbitrarily smooth initial state, arbitrarily many moments
vanish without the need for an apodizing function. For all of
these examples, however, \erf{support} still holds. This is
possible because in all of these examples $P_{\rm{wv}}(\wp)$ takes
on negative values. That is, the WWM involves a nonclassical
momentum transfer.

\subsection{Unbounded $\psi_i$}
\label{sec:unbdd}

For our first example we consider infinitely narrow slits centered
at $\pm s/2$. The minimal which way measurement functions,
$\{O_\pm(x)\}$, mentioned above are used.  Evaluating \erf{wvmt2}
then gives \beq \label{unbounded} P_{\rm{wv}}(\wp)=\frac12 \bigl[
\delta(\wp)+\frac{\sin(\wp{s}/2\hbar)}{\pi\wp} \bigr] \eeq This
distribution is plotted in Figure 2 along with $P_i(p)$ and
$P_f(p)$ for the same initial state and measurement operators.

\begin{figure}
\begin{center}
\includegraphics[width=0.48\textwidth]{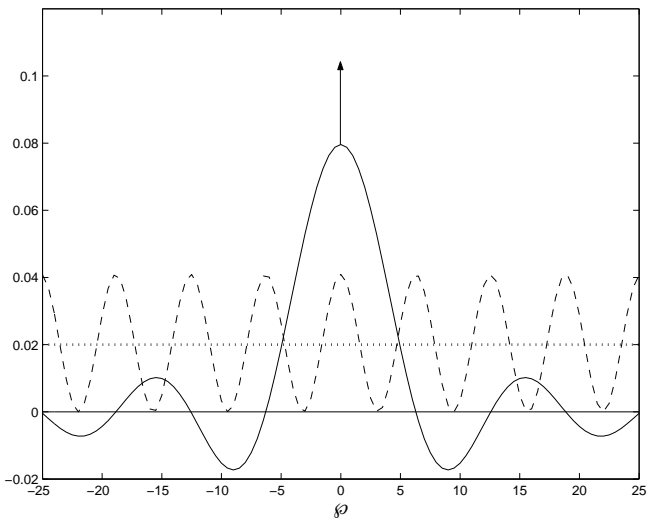}
\caption{Plots of the weak valued momentum transfer distribution,
$P_{\rm{wv}}(\wp)$, along with $P_i(p)$ and $P_f(p)$ for narrow
slits with $s=1$.  The solid line corresponds to
$P_{\rm{wv}}(\wp)$, the dashed line to $P_i(p)$ and the dotted
line to $P_f(p)$. The arrow denotes half of the delta function.
The momentum is scaled by taking $\hbar=1$.  It should be noted
that both the initial and final momentum distributions are in
reality spread out over all space in this case and are thus
actually infinitely small.  In order to compare them with
$P_{\rm{wv}}(\wp)$ they are shown here to be normalized over the
interval which they are plotted for.} \label{fig3}
\end{center}
\end{figure}

Obviously this is an example of a non-classical measurement as
$P_{\rm{wv}}(\wp)$ takes negative values periodically.  All
moments of order one and above are undefined in the strict sense.
If, however, we multiply by an apodizing function $f_\kappa(\wp)$
with characteristic width $\kappa$ that has all its moments
defined and smoothly goes to the unit function as
$\kappa\to\infty$, then we get the ``corrected" moments.  A simple
example is $f_\kappa(\wp)=\exp(-|\wp|/\kappa)$. We take the
apodized moments of $P_{\rm{wv}}(\wp)$ to be \bqa \label{apmom}
\langle{\wp^n}\rangle_{\rm{wv}} &=& \lim_{\kappa \to \infty} \int
d\wp P_{wv}(\wp)f_\kappa(\wp)\wp^{n} \eqa  For this example we get
\bqa \label{apmomeval} \langle{\wp^n}\rangle_{\rm{wv}} &=& \left.
\lim_{\kappa \to \infty}
\left[\frac{(1+(-1)^n)\Gamma(n)\sin(n\arctan(\kappa{s}/2\hbar))}{(1/2\kappa)^{n}
(4+(\kappa{s}/\hbar)^{2})^{n/2}} \right] \right.\nn\\&=& \left. 0
\,\,\forall{n}\right. \eqa

Even though the moments of this distribution are identically zero,
\erf{support} for its support is still satisfied. The
unit-confidence half-interval is found to be $[\delta \wp]_{1-{\rm
confidence}} \approx h/1.59s$. Thus even though the moments of
this distribution vanish, \erf{support} for the support of the
distribution is clearly satisfied. The non-zero width of $P_{\rm
wv}(\wp)$ can also be seen in its $1$-norm, as defined in
\erf{n-norm}. For this case it is necessary to use an apodization
function to calculate it.  We find that $[\delta \wp]_{1-{\rm
norm}} = 2h/\pi{s}$, which is again greater than $h/4s$.

\subsection{Bounded $\psi_i$}

We now consider another example, which is more realistic.
Consider the case of rectangular slits of width $w$ centered at
$\pm s/2$. We can write the wave function for this case as \bqa
\psi_i(x) &=&\left. \frac{1}{\sqrt{2w}}
\Theta(x+w/2)\Theta(-x+w/2) \ast (\delta(x+s/2)\right.\nn\\
&&\phantom{\frac{1}{\sqrt{2w}}}\left. + \delta(x-s/2)),\right.\eqa
where $f \ast g$ denotes the convolution of $f$ with $g$. In this
case we find that \beq \label{rectslits} P_{\rm{wv}}(\wp) =
\frac12[\delta(\wp) + \frac{2}{\pi w \wp^2}
\sin(\wp{s}/2\hbar)\sin(\wp{w}/2\hbar)]. \eeq

\begin{figure}
\begin{center}
\includegraphics[width=0.48\textwidth]{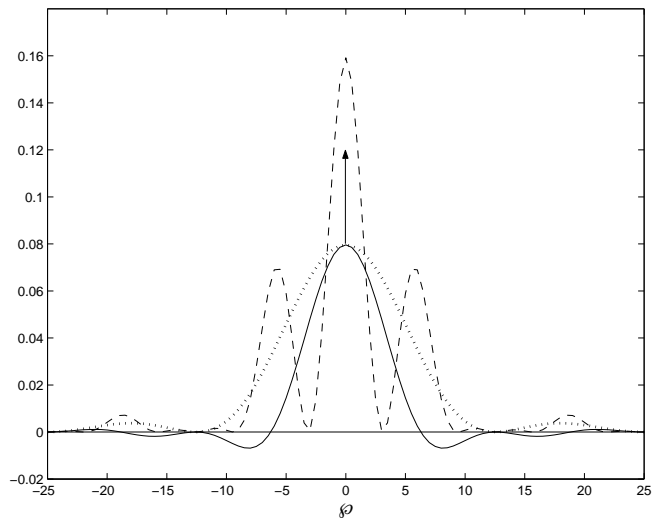}
\caption{Plots of the weak valued momentum transfer distribution,
$P_{\rm{wv}}(\wp)$, along with $P_i(p)$ and $P_f(p)$ for
rectangular slits with width $w=1/2$.  Other details are as in
Fig.~2, but note that here $P_i$ and $P_f$ are correctly
normalized.} \label{fig4}
\end{center}
\end{figure}

This distribution is plotted in Figure 3 along with $P_i(p)$ and
$P_f(p)$. This distribution obviously has its first moment defined
without need for apodization.  Second and higher order moments,
however, remain undefined in the strict sense.

\subsection{Continuous $\psi_i$}

For our next example, we consider a bounded and continuous initial
state.  Specifically, \bqa \psi_i(x) &=&\left.
\frac{1}{\sqrt{w}}\cos(\frac{\pi}{w}x)\Theta(x+w/2)\Theta(-x+w/2)
\right.\nn\\ &&\phantom{\frac{1}{\sqrt{w}}}\left. \ast
(\delta(x+s/2) + \delta(x-s/2)).\right.\eqa Given this initial
state and the minimal WWM, we find that \bqa \label{cosslits}
P_{\rm{wv}}(\wp) &=& \frac12\left[\delta(\wp) -
\frac{\sin(\wp{s}/2\hbar)}{\hbar{w}\wp}\left( \frac{\sin(\frac{w}{2\hbar}(\wp-\frac{\pi\hbar}{w}))}{\wp-\frac{\pi\hbar}{w}} \right.\right. \nn\\
&& \phantom{\frac12[} \left.\left. +
\,\frac{\sin(\frac{w}{2\hbar}(\wp+\frac{\pi\hbar}{w}))}{\wp+\frac{\pi\hbar}{w}}\right)\right] \\
&=&
\frac12\left[\delta(\wp)-\frac{2\pi\sin(\wp{s}/2\hbar)\cos(\wp{w}/2\hbar)}{(\hbar{w}\wp)[\wp^2-(\pi\hbar/w)^2]}\right].\nn\\
\eqa

\begin{figure}
\begin{center}
\includegraphics[width=0.48\textwidth]{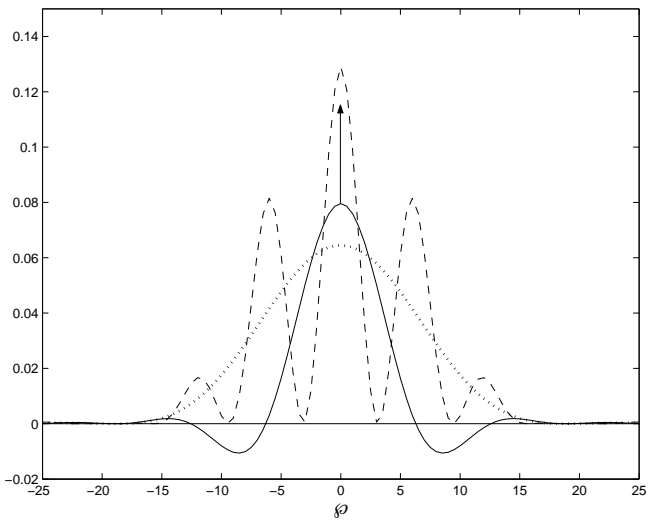}
\caption{Plots of the weak valued momentum transfer distribution,
$P_{\rm{wv}}(\wp)$, along with $P_i(p)$ and $P_f(p)$ for the
continuous initial wave function defined in Sec.~VI~C.  Other
details are as in Fig.~3.} \label{fig5}
\end{center}
\end{figure}

This distribution is plotted in Figure 4 along with $P_i(p)$ and
$P_f(p)$.  This distribution has both its first and second moments
defined without need for apodization. They vanish, of course.  The
theory of apodization is still required, however, to obtain the
correct higher order moments.

\subsection{Smooth $\psi_i$}

In the previous sections we found that more moments become defined
without the need for an apodizing function as we consider smoother
initial states. We will now consider what happens as our initial
state becomes differentiable to higher and higher order. That is,
we shall consider the case of \bqa \psi_i(x) &=&\left.
\frac{1}{\sqrt{N}}\cos^n\left(\frac{\pi}{w}x\right)\Theta(x+w/2)\Theta(-x+w/2)
\right.\nn\\ &&\phantom{\frac{1}{\sqrt{w}}}\left. \ast
(\delta(x+s/2) + \delta(x-s/2)),\right.\eqa  where \beq
N=\frac{4w}{\sqrt{\pi}}\frac{\Gamma(1/2+n)}{\Gamma(1+n)}.\nn\eeq
Note that the previous two examples are special cases of this,
corresponding to $n=0$ for Sec.~VI~C and $n=1$ for Sec.~VI~D. By
considering this general case we will be able to make our initial
state arbitrarily smooth by considering arbitrary degrees of $n$.
For the case of general $n$ we find \bqa \label{cosnslits}
P_{\rm{wv}}(\wp) &=& \frac12\left[\delta(\wp) +
\frac{1}{2^{n+1}\wp}\sin(\wp{s}/2\hbar)\sum_{k=0}^{n} {n \choose
k}
\right.\nn\\
&& \phantom{\frac12[}\left. \times
\frac{\sin(\frac{w}{2\hbar}(\wp-\frac{(n-2k)\hbar\pi}{w})}{\wp-\frac{(n-2k)\hbar\pi}{w}}\right].
 \eqa  For the case of even $n$ this simplifies to \bqa
\label{evenn} P_{\rm{wv}}(\wp) &=&
\frac12 \left[\delta(\wp)+\frac{n!}{2}\left(\frac{\pi\hbar}{w}\right)^n(-1)^{n/2}\right.\\
&&\left.\times\frac{\sin(\wp{s}/2\hbar)\sin(\wp{w}/2\hbar)}{(\wp-\frac{n\hbar\pi}{2})(\wp-\frac{(n-2)\hbar\pi}{2})\ldots(\wp+\frac{n\hbar\pi}{2})}\right].\nn\eqa
For odd $n$ it becomes \bqa \label{oddn} P_{\rm{wv}}(\wp) &=&
\frac12 \left[ \delta(\wp)+\frac{n!}{2}\left(\frac{\hbar\pi}{w}\right)^n(-1)^{(n+1)/2}\right.\\
&&\left.\times\frac{\sin(\wp{s}/2\hbar)\cos(\wp{w}/2\hbar)}{(\wp-\frac{n\hbar\pi}{2})(\wp-\frac{(n-2)\hbar\pi}{2})\ldots(\wp+\frac{n\hbar\pi}{2})}
\right].\nn\eqa

From \erf{evenn} and \erf{oddn}, for an initial state like the
cosine function raised to the power of $n$, it is easy to see that
$n+1$ moments are defined without the need for an apodizing
function. Thus as the initial state becomes smoother and smoother
arbitrarily many moments become defined without apodization.
Furthermore, all moments that are defined will be zero where the
measurement functions $O_\xi(x)$ are flat in the region of the
slits.

\section{Moments of $P_{\rm wv}(\wp)$ and other distributions}

In the preceding section we showed that in twin-slit schemes where
the measurement functions $O_\xi(x)$ are flat in the region of the
slits, all of the moments of $P_{\rm{wv}}(\wp)$ are zero. As
discussed in the introduction, these types of WWMs have no effect
on single slit diffraction patterns.  As such, the mean and
variance of a twin-slit diffraction pattern are not affected by
such a measurement.  Obviously then, the first and second moments
of $P_{\rm wv}(\wp)$ are equal to the change in the moments for
the momentum distribution of the particle.  In this section we
show that this is true for all WWMs, provided that the measurement
functions $O_\xi(x)$ vary slowly on the scale of the slits.
However, the third moment of $P_{\rm wv}(\wp)$ does {\em not}
correspond in general to the difference in the third moments of
the final and initial momentum distributions of the particle. We
show that this apparent discrepancy also arises from all known
formalisms investigating momentum transfer in WWMs (including a
new one introduced here). Moreover, with one exception, all of
these third moments are different.

\subsection{Moments of $P_{\rm wv}(\wp)$}

The moments of the weak valued momentum transfer probability
distribution, $P_{\rm{wv}}(\wp)$, can most easily be calculated
using the characteristic function of \erf{mgf}.  The $n^{\rm{th}}$
moment of this distribution is \bqa \label{nthwvmom}
\langle\wp^n\rangle_{\rm{wv}} &=& \left.
\left(-i\hbar\frac{\partial}{\partial{q}}\right)^n \sum_\xi \int
\!dx\, |\psi_{i}(x)|^{2} \half \bigl[
O_{\xi}(x)O_{\xi}^{*}(x-q)\right.\nn\\ && \phantom{\sum_\xi \int
\!dx\,} + O_\xi^*(x) O_\xi(x+q) \bigr]. \eqa  From this and the
completeness relation (\ref{complet}) it is easy to show that the
first three moments, for a completely arbitrary WWM, $O_{\xi}(x)$,
and initial state, $\psi_{i}(x)$, are \bqa \label{1mom}
\langle\wp\rangle_{\rm{wv}} &=& \left. -i\hbar \sum_\xi \int
\!dx\, |\psi_i(x)|^2 O_\xi^{*}(x)O_\xi'(x)\right. \\
\langle\wp^2\rangle_{\rm{wv}} &=& \left. \hbar^2 \sum_\xi \int
\!dx\, |\psi_i(x)|^2 {O_\xi^{*}}'(x)O_\xi'(x)\right. \label{2mom}\\
\langle\wp^3\rangle_{\rm{wv}} &=& \left. \frac{i\hbar^3}{2}
\sum_\xi \int \!dx\, |\psi_i(x)|^2
[O_\xi^{*}(x)O_\xi'''(x)\right.\nn\\
&& \phantom{\frac{i\hbar^3}{2} \sum_\xi \int }\left.
-O_\xi(x){O_\xi^{*}}'''(x)].\right. \label{3mom}\eqa

From this it is easy to see the results of the preceding section,
namely, that for the case where the measurement operators are
completely flat in the region of the slits, all moments of the
distribution vanish.

In order to more clearly compare the moments of the various
momentum transfer formalisms, it will be advantageous to consider
some special cases for $O_{\xi}(x)$ and $\psi_{i}(x)$. First we
consider the case of a weighted sum of unitary measurement
functions, that is \beq \label{unitaryO} O_\xi(x) = \sqrt{N_\xi}
\exp(i\phi_\xi(x)), \eeq where $\phi(x)$ is an arbitrary real
function.  \erfs{1mom}{3mom} then become \bqa
\label{wvunitarymom1} \langle\wp\rangle_{\rm{wv}} &=&\left.
\sum_\xi N_\xi \hbar \int \!dx\, {\phi_\xi'(x)} |\psi_i(x)|^2
\right.
\\\label{wvunitarymom2} \langle\wp^2\rangle_{\rm{wv}} &=&\left.
\sum_\xi N_\xi \hbar^2 \int \!dx\, {\phi_\xi'(x)}^2
|\psi_i(x)|^2 \right.\\
\langle\wp^3\rangle_{\rm{wv}} &=&\left. \sum_\xi N_\xi \hbar^3
\int \!dx\, [-\phi_\xi'''(x) +
{\phi_\xi'(x)}^3]|\psi_i(x)|^{2}.\right.\nn\\
\label{wvunitarymom3}\eqa

We now further consider the special case where the measurement
function varies slowly in the region of the slits. In this case we
can approximate the initial state by \beq \label{deltaslits}
\psi_i(x) = \sum_k \psi_k \varepsilon(x-x_k), \eeq where \beq
\label{normpsik} \sum_k |\psi_k|^2 = 1 \eeq and \beq
\label{epsilons} [\varepsilon(x)]^2 = \delta(x).\eeq  The
 moments are then just \bqa
\label{deltamom1}\langle\wp\rangle_{\rm{wv}} &=& \left. -i\hbar
\sum_{\xi,k} |\psi_k|^2 {O_\xi^{*}}(x_k)O_\xi'(x_k)\right.
\\\label{deltamom2} \langle\wp^2\rangle_{\rm{wv}} &=& \left.
\hbar^2 \sum_{\xi,k} |\psi_k|^2
{O_\xi^{*}}'(x_k)O_\xi'(x_k)\right.\\\label{deltamom3}
\langle\wp^3\rangle_{\rm{wv}} &=& \left. \frac{i\hbar^3}{2}
\sum_{\xi,k}
|\psi_k|^2 [O_\xi^{*}(x_k)O_\xi'''(x_k)\right.\nn\\
&& \phantom{\frac{i\hbar^3}{2} \sum_{\xi,k} |\psi_k|^2  }\left.
-O_\xi(x_k){O_\xi^{*}}'''(x_k)]\right.\eqa

Under both assumptions, the moments become \bqa \label{wv1special}
\langle\wp\rangle_{\rm{wv}} &=& \left. \hbar \sum_{k,\xi}
|\psi_k|^2 \phi_\xi'(x_k) \right.
\\\label{wv2special} \langle\wp^2\rangle_{\rm{wv}} &=& \left.
\hbar^2 \sum_{k,\xi} |\psi_k|^2 \phi_\xi'(x_k)^2 \right.\\
\langle\wp^3\rangle_{\rm{wv}} &=& \left. \hbar^3 \sum_{k,\xi}
|\psi_k|^2 [-\phi_\xi'''(x_k) + \phi_\xi'(x_k)^3].\right.\nn\\
\label{wv3special}\eqa

\subsection{Moments of $P_{f}(p)-P_i(p)$}

We now consider the differences in the moments of $P_{f}(p)$ and
$P_{i}(p)$.  That is, we want to look at \bqa
\langle{\hat{p}_f}^n\rangle - \langle{\hat{p}_i}^n\rangle
&=&\left. \sum_\xi
\bra{\psi_i}{\hat{O}_\xi}\dg\hat{p}^n\hat{O}_\xi\ket{\psi_i} -
\bra{\psi_i}\hat{p}^n\ket{\psi_i}\right.\nn\\\label{pf-pi}\eqa for
$n = 1,2,3$.

For the arbitrary measurement $O_\xi(x)$ and the initial state
$\psi_i(x)$ we find that the mean is identical to that found for
the weak valued distribution, \erf{1mom}. The second and third
moments are \bqa \langle{\hat{p}_f}^2\rangle -
\langle{\hat{p}_i}^2\rangle &=& \left. -\hbar^2 \sum_\xi \int
\!dx\,
[|\psi_i(x)|^{2}O_\xi^{*}(x)O_\xi''(x)\right.\nn\\
&& \phantom{-\hbar^2 \sum_\xi \int}\left. +
2\psi_i^{*}(x)\psi_i'(x)O_\xi^{*}(x)O_\xi'(x)]\right.\nn\\\label{2mompf-pi}\\
\langle{\hat{p}_f}^3\rangle - \langle{\hat{p}_i}^3\rangle &=&
\left. i\hbar^3 \sum_\xi \int \!dx\,
[|\psi_i(x)|^{2}O_\xi^{*}(x)O_\xi''(x)\right.\nn\\
&& \phantom{i\hbar^3 \sum_\xi \int}\left. +
3\psi_i^{*}(x)\psi_i'(x)O_\xi^{*}(x)O_\xi''(x)\right.\nn\\
&& \phantom{i\hbar^3 \sum_\xi \int}\left. +
3\psi_i^{*}(x)\psi_i''(x)O_\xi^{*}(x)O_\xi'(x)].\right.\nn\\\label{3mompf-pi}\eqa

In order to compare the second and third moments to those of the
preceding section,  we require the assumption that the measurement
functions vary slowly, i.e. we assume \erf{deltaslits}. Defining
the delta function in \erf{epsilons} to be \beq \label {defndelta}
\delta(x) = \lim_{\sigma \to 0} \frac{1}{\sqrt{2\pi{\sigma}^{2}}}
\exp(-x^{2}/2{\sigma}^2), \eeq one can easily show that \beq
\psi_i^{*}(x)\psi_i'(x) = \frac12 \sum_k |\psi_k|^2
{\delta}'(x-x_k) \eeq and \bqa \psi_i^{*}(x)\psi_i''(x) &=&\left.
\frac14 \lim_{\sigma \to 0} \sum_k |\psi_k|^2
[{\delta}''(x-x_k)\right.\nn\\&& \phantom{\frac14 \lim_{\sigma \to
0} \sum_k}\left. +\frac{1}{\sigma^2}\delta(x-x_k)].\right. \eqa
Using only the assumption that the measurement functions vary
slowly on the scale of the slits we find that the second moment is
identical to the analogous case for $P_{\rm{wv}}(\wp)$ as given in
\erf{deltamom2}. The third moment, however, is found to be \bqa
\label{3deltpi-pf} \langle{\hat{p}_f}^3\rangle -
\langle{\hat{p}_i}^3\rangle &=&\left. \frac{i\hbar^3}{4}
\lim_{\sigma \to 0} \sum_{\xi,k} |\psi_k|^2
[O_\xi^{*}(x_k)O_\xi'''(x_k)\right.\nn\\ && \left. +
3{O_\xi^{*}}''(x_k)O_\xi'(x_k)
-\frac{3}{\sigma^2}O_\xi^{*}(x_k)O_\xi'(x_k)].\right.
\nn\\\label{3deltpi-pf}\eqa  This looks nothing like the analogous
case for $P_{\rm{wv}}(\wp)$ and indeed diverges.  If we further
assume a weighted sum of unitary measurement operators this
becomes \bqa \label{3specialpi-pf} \langle{\hat{p}_f}^3\rangle -
\langle{\hat{p}_i}^3\rangle &=&\left.\hbar^3 \lim_{\sigma \to 0}
\sum_{k,\xi} N_\xi |\psi_k|^2[\phi_\xi'(x_k)^3 - \frac14\phi_\xi'''(x_k)\right.\nn\\
&&\phantom{\hbar^3 \lim_{\sigma \to 0} \sum_{k} |\psi_k|^2[}\left.
+ \frac{3}{4\sigma^2}\phi_\xi'(x_k)].\right. \eqa  Thus in general
the moments of $P_{\rm{wv}}(\wp)$ fail to have a relationship to
the moments of $P_{f}(p)-P_i(p)$ for second and higher order
moments. In the special case where the measurement function varies
slowly in the region of the slits, the second moments of
$P_{\rm{wv}}(\wp)$ and $P_{f}(p)-P_i(p)$ are identical.  For third
and higher order moments, however, there ceases to be any
relationship between these moments.  Indeed, \erf{3specialpi-pf}
may diverge.

\subsection{Moments of $\hat{p}_f - \hat{p}_i$}

An intuitively attractive measure of the momentum transfer in a
WWM that, to our knowledge, has not been used before is the
moments of the operator representing the difference between the
final momentum and the initial momentum.  This is defined in the
\hei\, picture.  That is, $\hat{p}_f - \hat{p}_i$ or alternatively
$$\hat{p}(\tau) - \hat{p}(0) =
\hat{U}\dg(\tau)\hat{p}\hat{U}(\tau) - \hat{p},$$ where here
$\hat{U}(\tau) = \hat U_{{}_{\rm WWM}}$ since free evolution
conserves momentum.  Following the notation introduced in Section
3, we find for the first moment \bqa \label{pfpi1}
\langle(\hat{p}_f-\hat{p}_i)\rangle &=&\left.
\bra{\psi_i}\mbra{\theta}(\hat U_{{}_{\rm WWM}}\dg\hat{p}\hat
U_{{}_{\rm WWM}} - \hat{p})\mket{\theta}\ket{\psi_i}\right.\nn\\
&=& \left. \sum_\xi\bra{\psi_i}\mbra{\theta}\hat U_{{}_{\rm
WWM}}\dg\mket{\xi}\hat{p}\mbra{\xi}\hat U_{{}_{\rm WWM}}
\mket{\theta}\ket{\psi_i}\right.\nn\\ && \phantom{\sum_\xi}\left. - \bra{\psi_i}\hat{p}\ket{\psi_i}\right.\nn\\
&=& \left.\sum_\xi
\bra{\psi_i}(\hat{O}_\xi\dg\hat{p}\hat{O}_\xi-\hat{p})\ket{\psi_i}.
\right. \eqa In the same way higher order moments are found to be
\bqa \label{pfpin} \langle(\hat{p}_f-\hat{p}_i)^n\rangle &=&\left.
\sum_\xi \bra{\psi_i}(\hat{O}_\xi\dg\hat{p}\hat{O}_\xi-\hat{p})^n
\ket{\psi_i}.\right. \eqa

If we evaluate these for arbitrary $\ket{\psi_i}$ and
$\hat{O}_\xi$ we find that the first two moments are identical to
those of the weak valued distribution in \erfs{1mom}{2mom}. The
third moment, however, is different from both previous cases
considered. It is \bqa \label{3mompfpi}
\langle(\hat{p}_f-\hat{p}_i)^3\rangle &=& \left. -i\hbar \int \!dx
|\psi_i(x)|^2O_\xi^{*}(x){O_\xi^{*}}'(x)\right.\nn\\
&& \phantom{-i\hbar \int \!dx} \left. \times
O_\xi'(x)^2\right.\eqa  If we now consider the two special cases
described above, we find that this is just \beq
\label{3mompfpispecial} \langle(\hat{p}_f-\hat{p}_i)^3\rangle =
\sum_{k,\xi} |\psi_k|^2 \phi_\xi'(x_k)^{3}. \eeq

\subsection{Moments of $P^{\rm Wigner}_{\rm local}(\wp)$}

We will now consider the Wigner function formalism introduced and
developed in \cite{Wis97a}. The Wigner function formalism has many
of the same properties as that of the weak value method.  In
particular, the local Wigner probability density for momentum
transfer, $P^{\rm Wigner}_{\rm local}(\wp)$, can take negative
values, just as the weak valued probability distribution.
Following Ref. \cite{Wis97a}, we have \beq \label{Wigdist} P^{\rm
Wigner}_{\rm local}(\wp) = \int \!dx |\psi_i(x)|^2 \sum_\xi
W_\xi(x,p),\eeq where $W_\xi$ is the Wigner function for
$O_\xi(x)$. The characteristic function for the Wigner function
probability distribution is given by \bqa \label{Wigchar} \Phi(q)
= \sum_\xi \int \!dx |\psi_i(x)|^2 O_\xi^{*}(x-q)O_\xi(x+q). \eqa
Using this it is easy to show that the first two moments are in
the completely general case identical to the those of the other
formalisms considered thus far (given in \erfs{1mom}{2mom}).
Again, the third moment is different from all the other cases
considered so far and is given by \bqa \label{Wigmom3}
\langle\wp^3\rangle_{\rm{local}}^{\rm{Wigner}} &=&\left.
\frac{i\hbar^3}{4} \sum_\xi \int \!dx |\psi_i(x)|^2
(O_\xi^{*}(x)O_\xi'''(x)\right.\nn\\ &&
\phantom{\frac{i\hbar^3}{4} \sum_\xi \int}\left. +
3{O_\xi^{*}}''(x)O_\xi'(x))\right. \eqa

For the special case of a slowly varying, unitary measurement
function we find for the third moment \bqa \label{Wigmom3special}
\langle\wp^3\rangle_{\rm{local}}^{\rm{Wigner}} &=&\left.
\sum_{k,\xi} |\psi_k|^2(\phi_\xi'(x_k)^3 -
\frac14\phi_\xi'''(x_k)).\right. \eqa

\subsection{Moments of $P^{\rm Bohm}_{\rm local}(\wp)$}

Finally, we consider the Bohmian formalism, as introduced in
\cite{Wis98a}. In this formalism the particles in a WWM have a
definite position $x$ and momentum $p = m\dot{x} =
\rm{Re}[-i\hbar\psi'(x)/\psi(x)]$.  The probability distribution
for $x$ is as usual, $|\psi(x)|^2$ but the probability
distribution for $p$ is in general not simply
$|\tilde{\psi}(p)|^2$ and only becomes this in the far field.
Because particles have a definite position and momentum in Bohmian
mechanics, it is possible to track their trajectories and in turn
calculate a time-dependent momentum transfer probability
distribution, $P^{\rm{Bohm}}(\wp;t)$, where $t$ is the time after
the WWM. In this formalism the momentum continues to change well
after the WWM, and thus does so in a non-local way.  The local
momentum transfer is given by $P^{\rm{Bohm}}(\wp;0^+)$.

In this formalism, the momentum transfer is in general critically
sensitive to the slit width, $w$, as discussed in detail in
\cite{Wis98a} . Because of this difficulty, it is best to consider
only the measurement functions that do not localize a particle at
one slit.  That is, measurement functions of the form given in
\erf{unitaryO}. In this case, the local Bohmian momentum transfer
distribution is given by \bqa \label{bohmdist}
P_{\rm{local}}^{\rm{Bohm}}(\wp) &=& \left. \sum_\xi N_\xi \int
\!dx |\psi_i(x)|^2\delta(\wp - \hbar\phi_\xi'(x)).\right.\nn\\
\eqa

From this equation it is trivial to find the moments.  The first
two are identical to those of the other formalisms, given in
\erfs{wvunitarymom1}{wvunitarymom2}.  The third moment,
interestingly, is the same as that of the moments of $\hat{p}_f -
\hat{p}_i$, that is \beq \label{bohmmom3}
\langle\wp^3\rangle_{\rm{local}}^{\rm{Bohm}} = \hbar^3\sum_\xi
N_\xi \int \!dx |\psi_i(x)|^2 \phi_\xi'(x)^3. \eeq

\subsection{Comparison}

In summary, all formalisms known to us for quantifying the
momentum transfer in WWMs give the same mean and variance of the
momentum transfer for measurement functions that vary slowly in
the region of the slits.  With one exception, they all have
different third moments.  Moreover, none of these third moments
are equal to the difference in the third moments of $P_f(p)$ and
$P_i(p)$ (unlike the mean and variance). The fact that only the
mean and variance  of $P_{\rm{wv}}(\wp)$ are relevant to the
change in the moments of the momentum distributions thus is not
unique to this formalism.
 Since the first and second moments are the most important ones
  for characterizing a probability
distribution, our conclusion is that there is no point considering
 higher order moments.

\section{Conclusion}

The question of whether which-way measurements destroy
interference by disturbing the momentum of the particle has been
the subject of debate for a  decade now. What has been lacking has
been a way to address this question in a meaningful and
interesting way experimentally. In this paper we have, following
Ref.~\cite{Wis03}, analysed the applicability of weak values to
the question. Unlike other formalisms, the concept of weak values
allows a pseudo-probability distribution $P_{\rm wv}(\wp)$ for
momentum transfer to be directly observed experimentally.

The distribution $P_{\rm wv}(\wp)$ has many attractive features.
It depends on the initial state of the particle and on the
functions that define the WWM, in a very simple and intuitive way.
It reproduces the distribution $P_{\rm cl}(\wp)$ in the case where
the momentum transfer is classical (and therefore unambiguous).
The nicest feature is that the single function, $P_{\rm wv}(\wp)$,
is compatible with both sides of the debate. In support of Storey,
Tan, Collett and Walls, it can be shown that the (suitably
defined) width of $P_{\rm wv}(\wp)$ is always at least $h/6s$.
However, in support of Scully, Englert and Walther, for the WWM
they proposed  the mean and variance of $P_{\rm wv}(p)$ are zero.
This is true for all WWMs where the measurement functions are
completely flat in the region of the slits and thus leave a single
slit diffraction pattern unchanged.

Furthermore, we have shown here that for a general WWM where the
measurement functions vary slowly on the scale of the slits, the
mean and variance of $P_{\rm wv}(\wp)$ are equal to the change in
the moments of $P_f(p)$ and $P_i(p)$.  This relationship does not
hold for higher order moments, but nor does it hold for any of the
formalisms that have been developed thus far for quantifying
momentum transfer in WWMs.
 In fact, all such formalisms (with
one exception) yield different third moments, while they all yield
the same mean and variance, that of the weak valued distribution.

The attractive feature that $P_{\rm wv}(\wp)$ agrees with both SEW
and STCW is possible only because it is a pseudo-probability
distribution: it can can take negative values. This does not mean
something is wrong with the formalism. It must be remembered that
$P_{\rm wv}(\wp)$ is {\em not} obtained from an experiment as the
relative frequency of measuring a momentum transfer of $\wp$.
Rather, it is itself the average of (weak) measurement results.
The important point is that in an experiment $P_{\rm wv}(\wp)$ can
be inferred from measurement results using only classical
reasoning. That is, a classical physicist would expect $P_{\rm
wv}(\wp)$ to be a true probability distribution.

This is exactly analogous to the reconstruction of $W(x,p)$ by
homodyne tomography. The {\em process} is ``fully understandable
classically", to use the words of Alexander Lvovsky at ICSSUR03 \cite{Lvo03}.
Only the {\em product} [$W(x,p)$, or, in the present case, $P_{\rm
wv}(\wp)$] is classically impossible. Not only is $P_{\rm
wv}(\wp)$ measurable in principle; the techniques used in the
first weak-valued experiment \cite{RitStoHul91} are readily
adaptable to the twin-slit situation considered in this paper.
Thus interesting instances of this distribution, such as that in
\erf{rectslits}, should soon be subject to experimental
verification.

\section*{Acknowledgments}
This work was supported by the Australian Research Council.

\begin{appendix}
\section{The Mathematical Formalism of Weak Values}

Consider a family ($0<\sigma <\infty$) of measurements of an
observable $\hat{X}$ with the probability distribution for the
results $x$ being $P_\sigma(x) =
\bra{\psi}\hat{E}_\sigma(x)\ket{\psi}$ where \beq
\hat{E}_{\sigma}(x) = (2\pi\sigma^{2})^{-1/2}
\exp\cu{-(x-\hat{X})^{2}/2\sigma^{2}} \eeq is called the {\em
probability operator} for the measurement. In the limit $\sigma
\to 0$ one gets {\em strong} or precise measurements: $E_\sigma(x)
\to \ket{x}\bra{x}$. In the opposite limit, where $\sigma$ is arbitrarily large,
one gets {\em weak} measurements. Note that
as $\sigma \to \infty$, $\hat E_\sigma(x) \to \hat{1}$, the identity
operator, which leaves the system completely unchanged.

For a minimally-disturbing measurement, the measurement operator
is simply the square root of the probability operator. In other
words, the {\em conditional} system state given the result $x$ is
\beq
 \ket{\psi_{x}} = {\hat{E}^{1/2}_{\sigma}(x)} \ket{\psi}/\sqrt{P_\sigma(x)}.
 \eeq
If the system evolves unitarily after the weak measurement, the
probability for it to be found in state $\ket{\phi}$, given the
result $x$, is thus \beq
  P_{\sigma}(\phi|x) =
\st{ \bra{\phi}\hat{U}{\hat{E}^{1/2}_{\sigma}(x)}
\ket{\psi}}^{2}/P_{\sigma}(x).
  \eeq

Now using Bayes' theorem we have
 \beq
 P_\sigma(x|\phi) = P_{\sigma}(\phi|x) P_\sigma(x)\,/\int dx P_{\sigma}(\phi|x)P_\sigma(x).
 \eeq
From this, with a little care, it can be shown that
\beq
 \lim_{\sigma\to\infty} \int dx\,x\, P_{\sigma}(x|\phi)  = {\rm
Re}\frac{\bra{\phi}\hat U\hat{X}\ket{\psi}}{\bra{\phi}\hat
U\ket{\psi}},
 \eeq
which is the result quoted in \erf{weakvalgen}.

\end{appendix}


\begin{references}
\bibitem{fn1}
As in the previously stated result for the disturbance of a
momentum eigenstate, the width of $P_{\rm wv}(\wp)$ is actually
related only to the quality $Q$ \cite{MarHar03} of the WWM, not
the visibility $V$ (which equals $Q$ for a symmetric twin-slit
initial state). That is, this property of $P_{\rm wv}(\wp)$ still
holds even if $\ket{\psi_i}$ is a {\em single-slit} initial state.
\end{references}
\end{document}